\documentclass[showpacs,preprintnumbers,amsmath,amssymb]{revtex4}

\usepackage{graphicx}% Include figure files
\usepackage{dcolumn}% Align table columns on decimal point
\usepackage{bm}% bold math

\newcommand{\be}{\begin{equation}}
\newcommand{\ee}{\end{equation}}
\newcommand{\ben}{\begin{equation*}}
\newcommand{\een}{\end{equation*}}
\newcommand{\bea}{\begin{eqnarray}}
\newcommand{\eea}{\end{eqnarray}}
\newcommand\re[1]{(\ref{#1})}

\newcommand{\eqnsection}{
\renewcommand{\theequation}{{\thesection.\arabic{equation}}}
\makeatletter \csname @addtoreset\endcsname{equation}{section}
\makeatother} \eqnsection

\begin{document}

\title{On the evolution of higher order fluxes
in non-equilibrium thermodynamics}

\author{V. Ciancio*, V. A. Cimmelli{**} and P.
V\'an{***}}

\address{
*Department    of     Mathematics,     University of Messina,
Contrada Papardo, Salita Sperone, 31, 98166, Messina, Italy \\
E-mail:ciancio@dipmat.unime.it \\
**Department of Mathematics, University of
Basilicata, Campus Macchia Romana, 85100 Potenza, Italy \\
E-mail:cimmelli@unibas.it \\
***Department of Chemical Physics, Budapest University of
Technology and Economics, Budafoki \'ut 8, 1521 Budapest, Hungary \\
E-mail:vpet@eik.bme.hu }


\begin{abstract}
The connection between the balance structure of the evolution
equations of higher order fluxes and different forms of the entropy
current is investigated on the example of rigid heat conductors.
Compatibility conditions of the theories are given. Thermodynamic
closure relations are derived.
\medskip

\end{abstract}

\maketitle

\markboth{\small\scshape V. Ciancio, V. A. Cimmelli, P. V\'an
}{\small\scshape Balance Laws and Entropy Fluxes...}

\section{Introduction}

In the last two decades two different thermodynamic theories have
been applied in modeling fast non-equilibrium phenomena: the
Extended Thermodynamics \cite{MulRug98b,JouAta92b,JouAta01a} and the
Irreversible Thermodynamics with dynamic variables
\cite{Ver97b,MusAta01a}. In the following investigations
instead of referring directly to persons we sometimes distinct
theories of Extended Thermodynamics (Rational Extended
Thermodynamics \cite{MulRug98b}, Extended Irreversible
Thermodynamics \cite{JouAta92b,JouAta01a} and Wave Approach of
Thermodynamics \cite{Gya77a,MarGam89a}).

Extended Thermodynamics (ET) lies on the kinetic theory of gases
and postulates suitable balance equations for dissipative fluxes.
In this way a hierarchical system  in which the flux at step $n$
becomes the wanted field at step $n+1$, may be obtained.
Furthermore, the evolutionary systems are required to be symmetric
hyperbolic so that finite speed of propagation of disturbances and
well posedness of the Cauchy problem are guaranteed
\cite{MulRug98b}.

Irreversible Thermodynamics (IT) enlarges the basic state space by
introducing certain additional dynamic variables whose evolution is
ruled either by ordinary differential equations (local theory)
\cite{Ver97b,KluCia78a} or by partial differential equations (weakly
nonlocal theory) \cite{CimRog01a,Van03a}. In such a framework the
parabolic models are also allowed in that, once their solutions are
interpreted in the light of the experimental measurements, they also
lead to finite speed of propagation \cite{Cim04a}. ET is a special
case of IT with internal variables, where the form of the evolution
equations is restricted and the role of the internal variables is
fixed.

In the present paper we compare these two approaches by deriving
conditions under which they lead to the same system of equations.
To achieve that task we apply the classical Onsager analysis of
Second Law \cite{Ver97b,GroMaz62b}.

We point out the central role of the entropy current in the two
theories. In Rational Extended Thermodynamics the entropy current
is a derived quantity. Its form is restricted by the balance
structure of the evolution equations through the Second Law (see
e.g. \cite{MulRug98b}, page 24 or page 60).  The restriction can
be expressed by the following differential expression:
\begin{equation}
 {\rm d}j_i = \frac{\partial s}{\partial a} {\rm d}j^a_{\;\;i}
 \label{jcurrret}\end{equation}

Here $j_i$ is the $i$th component of the entropy current, $s$ is
the entropy density, $a$ and $j^a$ denotes the densities their
currents in the corresponding balances $\dot{a} + j^a_{i,i}
=0$,respectively. Here the entropy current appears as a potential
on the space of the currents of the extensive quantities.
Considering that the currents of the extensives $j^a$ are also
constitutive quantities, the above expression constraints
considerably their form. The surprisingly restrictive constraints
were proved to be compatible with some kinetic theory models. As a
price some of the higher order currents were considered as
constitutive quantities instead of being variables.

On the other hand in IT with dynamic variables \cite{Ver97b} the
starting point is the following form of the entropy current
\begin{equation}
j_i = \frac{\partial s}{\partial a} j^a_{\;\;i}.
 \label{jcurrnet}\end{equation}

This form was suggested as a direct generalization of the
classical heat flux over temperature expression $j_i = q_i/T$
\cite{Ver83a}.

The entropy current \re{jcurrret} is the consequence of the
balance structure of the evolution equations. On the other hand
from any given form of the entropy current one can determine, or
at least restrict, the form of the evolution equations of the
dynamic variables. In Extended Irreversible Thermodynamics one can
meet both forms, Jou, Casas-V\'asquez and Lebon seem to hesitate
on the advantages and disadvantages of the two suggestions
(compare \cite{DedAta96a}, \cite{JouAta92b} p. 138 and
\cite{JouAta92b}, p. 56-58, \cite{JouAta04a}).

In the following we will investigate the compatibility of the
Verh\'as form of the entropy current with the balance structure of
the evolution equations in a general and highly nonlinear case. We
give conditions when the balance structure and the \re{jcurrnet}
entropy current are compatible. Our investigations show that if we
do not insist strictly to the balance structure then the momentum
hierarchy can be closed easily on the phenomenological level.

In section 2, after considering a rigid body at rest and out of
local equilibrium, we specify the basic balance equations together
with the main constitutive assumptions. In section 3 we obtain the
evolution equations for the dynamic variables. In section 4 the
conditions are pointed out under which these equations yield a 4
moments model and a 13 moments model of rigid heat conductor of
Extended Thermodynamics. In section 5 we consider the propagation
of heat waves at low temperatures. We reinspect some extended
thermodynamic models \cite{DreStr93a} and determine the conditions
under which they can be recovered in the framework of the
presented theory. A concluding discussion of the obtained results
is developed in section 6.

\section{Balance equations and Second Law of thermodynamics}

As a simple but representative example let us consider a rigid
heat conductor at rest and let us start from the following local
balance equation of the internal energy \be \label{balinte}
 \dot{e} + q_{i'i} =0,
\ee \noindent where $e$ is the density of internal energy, $q_i$
$i=1,2,3$ are the components of the heat flux {\bf q}, $\dot{f}
\equiv \frac{\partial f}{\partial t}$, $f_{'i} \equiv
\frac{\partial f}{\partial x_i}$, $x_i$ $i = 1,2,3$ are the
Cartesian coordinates of the points of the body and the Einstein
convention of summation over the repeated indices has been
applied. The local balance of entropy is given by
\be\label{balent}
 \dot{s} + j_{i'i} =\sigma_s,
\ee

\noindent with $s$ standing for the entropy density,  $j_i$
$i=1,2,3$ for the components of the entropy current ${\bf j}$ and
$\sigma_s$ for the density of entropy production. Second Law of
Thermodynamics forces $\sigma_s$ to be nonnegative. The only
equilibrium variable will be the internal energy $e$, while the
first dynamic variable is supposed to be the heat flux itself.
Such an assumption, following by the celebrated Cattaneo's
equation \cite{Cat48a}, is included within the basic postulates of
both theories under examination (see \cite{MulRug98b} Chapter 1,
and \cite{Ver97b}). Hence, it also will be our basic starting
point. As a further dynamic variable let us choose a second order
tensor $\boldsymbol{\Phi}$ whose components will be denoted by
$\Phi_{ij}$, $i,j=1,2,3$. The physical meaning of $\Phi_{ij}$ will
be specified in section 5 on the example of second sound wave
propagation.\\ Hence we postulate the constitutive equation as
\be\label{locconst}
 {F}={F}^*(e,q_i,\Phi_{ij}),
\ee

\noindent where ${F}$ means an element of the set of constitutive
functions $\{s, {\bf j}\}$. The basic thermodynamic state space is
spanned by the 13 unknown quantities $\{e,q_i,\Phi_{ij}\}$ but
later on some enlargements will be considered. Beside
Eq.\re{balinte}, the system of evolution equations for the 12
wanted fields $q_i$ and $\Phi_{ij}$ is needed. In order to derive
such a system let us calculate $\sigma_s$ according to the
classical procedures of Irreversible Thermodynamics. Along with
Gyarmati and Verh\'as \cite{Gya77a,Ver83a} we represent the
entropy function, out of local equilibrium, as
\be\label{Gyarent}
s(e,q_i,\Phi_{ij}) = s_0(e) + \frac{1}{2} m_{ij} q_i
    q_j + \frac{1}{2} n_{ijkl} \Phi_{ij} \Phi_{kl}.
\ee

In \re{Gyarent} $s_0(e)$ is the local equilibrium entropy while the
matrices of thermodynamic inductivities $m_{ij}$ and $n_{ijkl}$ are
constitutive functions depending on the basic fields $e$, $q_i$ and
$\Phi_{ij}$. Moreover, we assume that the local equilibrium is
globally thermodynamically stable and there are no phase boundaries
at the non-equilibrium part of the state space. Therefore, due to
the principle of maximum  entropy  at the equilibrium, $m_{ij}$ and
$n_{ijkl}$ are negative definite. As the antisymmetric part of the
inductivities {\bf m} and {\bf n} do not contribute to the entropy
we assume that they are symmetric as follows:
\be
m_{ij} = m_{ji}, \qquad n_{ijkl}=n_{klij}.
\label{Sim1}\ee
Finally, along with Verh\'as \cite{Ver83a,Ver96a,DedAta96a} we
represent the entropy current as
\be\label{Verecurr}
  j_i = \frac{\partial s}{\partial e} q_i +
  \frac{\partial s}{\partial q_j} \Phi_{ji}.
\ee

Let us observe that the previous form of the entropy current is
not the most general one and some extensions are also possible
(see for instance \cite{Nyi91a1}). Due to \re{Gyarent}, $j_i$
specialize to
\begin{eqnarray}\label{entcurr}
j_i &=& \frac{\partial s_0}{\partial e}q_i + \frac{1}{2} A_{hk}q_h
q_k q_i +
  \frac{1}{2} B_{hklm}\Phi_{hk} \Phi_{lm}q_i + \\
   \frac{1}{2} C_{hkj}q_h q_k \Phi_{ji} +
   &+&  m_{hj}q_h \Phi_{ji} +
  \frac{1}{2} D_{hklmj}\Phi_{hk} \Phi_{lm} \Phi_{ji}. \nonumber
\end{eqnarray}

and
$$
A_{hk} \equiv \frac{\partial m_{hk}}{\partial e}; \quad B_{hklm}
\equiv \frac{\partial n_{hklm}}{\partial e}; \quad C_{hkj} \equiv
\frac{\partial m_{hk}}{\partial q_j}; \quad D_{hklmj} \equiv
\frac{\partial n_{hklm}}{\partial q_j}.
$$

In the next section we will use \re{Gyarent} and \re{entcurr} in
order to calculate the most important quantity related to this
procedure, namely, the density of entropy production. Let us
observe the special dependence of  vector {\bf j} on the
components $q_i$ and $\Phi_{ij}$ of the basic field. Due to
\re{entcurr}  the scalar $j_{i'i}$ does not contain the divergence
of any third order tensor to be interpreted as the flux of
$\boldsymbol{\Phi}$. As far as the form of the evolution equations
is concerned, such a property results to be crucial.

\section{Bilinear form of the entropy production}

Our aim here is to prove that the entropy production $\sigma_s$
can be put in the classical bilinear form of Irreversible
Thermodynamics \cite{GroMaz62b}. Some straightforward calculations
show that, due to \re{Gyarent} and \re{entcurr} the function
$\sigma_s$ can be written as follows
\be
\sigma_s =
   \left[\frac{\partial^2 s}{\partial e^2}e_{'i} +
   \frac{1}{2} A_{ik} q_{l'l} q_k +
    A_{hk} q_h q_{k'i} +
    \frac{1}{2} A_{hk'i} q_h q_k + \right. \nonumber\\
\ee
\be
    + \frac{1}{2} B_{hklm'i} \Phi_{hk} \Phi_{lm} +
    \frac{1}{2} (B_{hklm}+B_{lmhk}) \Phi_{hk'i} \Phi_{lm} +
    \frac{1}{2} C_{ikl'm} q_k \Phi_{lm} + \nonumber \\
\ee
\be    + C_{hil} q_{h'm} \Phi_{lm} +
    \frac{1}{2} C_{hil} q_h \Phi_{lm'm} +
    m_{hi'l} \Phi_{hl} + m_{hi} \Phi_{hl'l} + \nonumber\\
\ee
\be
    + \left. \frac{1}{2} A_{ij} q_j \dot{e} +
    \frac{1}{2} C_{ijk} q_j \dot{q_k} +
    m_{[ik]} \dot{q_k} +
    \frac{1}{2}F_{ijkl}\dot \Phi_{kl} q_{j}\right] q_i +
  \left[\frac{1}{2} B_{hklm} \Phi_{lm}q_{i'i} +
  \right.\nonumber \\
\ee
\be
   + m_{hp} q_{p'k} +  \frac{1}{2} D_{hklmp'i} \Phi_{lm}\Phi_{pi} +
    D_{hklmp}\Phi_{lm'i}\Phi_{pi} + \nonumber\\
\ee
\be    + \frac{1}{2} D_{hklmp} \Phi_{lm}\Phi_{pi'i}
    + \frac{1}{2} B_{hklj} \Phi_{lj}\dot{e}
     + \frac{1}{2} D_{hkjlp} \dot{q_p}\Phi_{jl} + \nonumber\\
\ee
\be    + \frac{1}{2} G_{hkjlpq} \Phi_{jl}\dot\Phi_{pq} +
     \left. n_{hklm}\dot \Phi_{lm}
   \right]\Phi_{hk}.
\label{longform}
\ee

In deriving \re{longform} we have put
$$
F_{ijkl} \equiv \frac{\partial m_{ij}}{\partial \Phi_{kl}}; \qquad
G_{ijklpq} \equiv \frac{\partial n_{ijkl}}{\partial \Phi_{pq}}.
$$

A direct inspection of \re{longform} shows that each term stands
for the combination of a relaxation and a transport. The
quantities in the square brackets are regarded as process forces
while their coefficients $q_i$ and $\Phi_{ij}$ as thermodynamic
fluxes. Because the expression \re{longform} of the entropy
production is of the usual bilinear form, let us apply Onsager's
linear equations. In this way we get the evolution equations we
are looking for, i. e.
\begin{eqnarray}
[{\bf q}]_i &=& L_{ik} q_k + M_{ikl}\Phi_{kl} \label{oc1}\\
\left[\boldsymbol{\Phi}\right]_{hk} &=& P_{hkl}q_l +
Q_{hkij}\Phi_{ij} \label{oc2}\end{eqnarray}

\noindent where the symbols $[{\bf q}]_i$ and
$[\boldsymbol{\Phi}]_{hk}$ stand for the coefficients of $q_i$ and
$\Phi_{hk}$ respectively and the tensor functions  ${\bf L}$,
${\bf M}$$, {\bf P}$, ${\bf Q}$ are constitutive functions
depending on the basic fields. With these conditions the above
form is the solution of the entropy inequality.

It is important to remark here, that the classical Coleman-Noll
\cite{ColNol63a} and Liu \cite{Liu72a} procedures for the
exploitation of Second Law are not suitable to obtain the equations
above since these techniques lead to thermodynamic restrictions on
the constitutive equations which are not in the form of
phenomenological evolution equations. After applying these
exploitation techniques one can get a restricted form of the
dissipation inequality. The Onsagerian approach could start here,
with the identification of the fluxes and the forces. Indeed, the
essence of the Onsager approach is the strict distinction between
the undetermined constitutive quantities (currents) and given
functions of the constitutive space (forces). With a suitable
distinction and some continuity requirements one can give the {\em
general solution} of the dissipation inequality in the form of
dynamic equations \cite{Gur96a}.

Here, in this paper we have chosen more direct approach with
assuming suitable specific forms of the entropy \re{Gyarent} and the
entropy current \re{Verecurr}. The resulted entropy production makes
possible to derive explicit evolution equations for the fluxes. The
procedure allows to close the system and this is due to the Second
Law.

The expressions \re{oc1} and \re{oc2} represent a system of 12
differential equations which, together with the balance of energy
\re{balinte}, allows - in principle - the determination of the 13
unknown functions $e$, $q_i$ and $\Phi_{ij}$. However it is not in
the balance form and hence, to achieve that task, some additional
assumptions seem to be necessary. This subject will be
investigated next.

\section{Balance form for evolutionary systems}

Up to now we did not restrict the inductivity tensors ${\bf m}$
and ${\bf n}$ at all. Let us assume now, that they are constant
and non-degenerate. Let us remark, that non-degeneracy is not
trivial, since there exist real materials for which it is not
guaranteed. A classical example is given in \cite{Mei73a1}, where
an electric circuit described by dynamic variables is considered.

For the sake of simplicity, let us assume that the inductivities
take the form
\be
 \label{simpl} {\bf m}=-m {\bf I},\,\,\,\,  {\bf n} = -n {\bf I},
\ee
where $m$ and $n$ are positive real coefficients and ${\bf I}$
means the unitary tensor in the corresponding tensorial space.
Finally, according to arguments from the kinetic theory of rigid
heat conductors, we assume that $\Phi$ is symmetric
\cite{MulRug98b,JouAta92b,DreStr93a}. Under the simplification
above the equations \re{oc1} and \re{oc2} yield
\begin{eqnarray}
-\dot{q_i} - \Phi_{ik'k} &=&
    \frac{1}{m}\Big(L_{ik} q_k + M_{ikl} \Phi_{kl} -
\frac{\partial^2 s_{0}}{\partial e^2}e_{'i} \Big), \label{linoc1}\\
-\dot\Phi_{ij} &=& \frac{1}{n}\Big(P_{ijk} q_k + Q_{ijkl}
\Phi_{kl} \Big) + \frac{m}{n} q_{i'j}. \label{linoc2}
\end{eqnarray}

Here we denoted the symmetries of the constitutive functions, too.
From \re{linoc1}-\re{linoc2} results in that the evolution of {\bf
q} is ruled by a set of balance laws with gradient dependent
source terms while tensor $\boldsymbol{\Phi}$ obeys a set of
ordinary differential equations (with a given {\bf q}). The system
\re{balinte}, \re{linoc1} and \re{linoc2} may be regarded as the
four moments theory for rigid heat conductors of Extended
Thermodynamics \cite{DreStr93a} provided the equations \re{linoc2}
are interpreted as suitable closure conditions
\cite{MulRug98b,JouAta92b}. This point of view is quite different
from that often applied in the current literature, where the
closure relations are deduced either by different assumptions
inside the kinetic theory \cite{MulRug98b,Lib90b}, by a maximum
entropy formalism \cite{Pap04a} or by some algebraic
considerations \cite{Edw02a}. To our opinion it is remarkable that
closure relations have been obtained in a fully thermodynamic
framework.

Let us prove now that the 10 moments theory for rigid heat
conductors may be obtained by a model with 27 dynamic variables.
Therefore in the following we assume that the introduced dynamic
variables are symmetric. In this way we can compare our results
with those of the kinetic theory. We assume the constitutive equation
\re{locconst} takes the form
\be\label{locconst1}
 {F}={F}^*(e,q_i,\Phi_{[ij]},\Psi_{[ijk]}),
\ee

\noindent where $\Phi_{[ij]}$ is a symmetric second order tensor and
$\Psi_{[ijk]}$ denotes a component of a fully symmetric third order
tensor $\boldsymbol{\Psi}$ which enters the basic field. Here and in
the following the bracket $_{[ij]}$ denotes a fully symmetric tensor
in the bracketed indices. As a consequence we can write the entropy
density as follows

\be\label{Gyarent1}
s(e,q_i,\Phi_{[ij]},\Psi_{[ijk]}) = s_0(e) + \frac{1}{2} m_{ij} q_i
q_j + \frac{1}{2} n_{ijkl} \Phi_{[ij]} \Phi_{[kl]} + \frac{1}{2}
r_{ijklpq} \Psi_{[ijk]}\Psi_{[lpq]},
\ee

\noindent where the meaning of the negative definite tensor ${\bf
r}$ is plausible. Here, in addition to the symmetry properties
\re{Sim1} we can see, that $n_{ijkl} = n_{jikl}$ and $n_{ijkl} =
n_{ijlk}$ that we could denote as $n_{[[ij][kl]]}$. Similarly, the
symmetry properties of {\bf r} could be denoted by
$r_{[[ijk][lpq]]}$. In the following we omit the notation of the
symmetries {\bf m}, {\bf n} and {\bf r}. Moreover, if ${\bf m}$,
${\bf n}$ and ${\bf r}$ are constant, we get
\be\label{Verecurr1}
j_i = \frac{\partial s_{0}}{\partial e} q_i +
 \frac{\partial s}{\partial q_{k}} \Phi_{[ki]} +
  \frac{\partial s}{\partial \Phi_{[kl]}} \Psi_{[kli]}=
 \frac{\partial s_{0}}{\partial e} q_{i} + m_{hk} \Phi_{[hi]}q_k
 + n_{ljhk} \Phi_{[hk]}\Psi_{[lji]},
\ee
Due to \re{Gyarent1} and \re{Verecurr1} the entropy production
takes the form
\begin{eqnarray}\nonumber
\sigma_s &=&
  \left[\frac{\partial^2 s_{0}}{\partial e^2}e_{'i} +
    m_{hi} \Phi_{[hj]'j} +
   m_{ik} \dot{q_k}  \right]q_i + \\ \nonumber
  &+& \left[m_{hp} q_{p'k} +  n_{hklm}{\Psi}_{[lmp]'p} +
    n_{hklm}\dot{\Phi}_{[lm]} \right]\Phi_{[hk]} + \\
  &+& \left[n_{ijhm}\Phi_{[hm]'k} + r_{ijklpq} \dot{\Psi}_{[lpq]}
   \right]\Psi_{[ijk]}.
\label{longform1}\end{eqnarray}

Again, due to (\ref{Gyarent1}) the matrices ${\bf m}$, ${\bf
n}$ and ${\bf r}$ are symmetric (and hence invertible) and we are
allowed to suppose that ${\bf m}$ and ${\bf n}$
take the form  (\ref{simpl}) while
\be
{\bf r} = -r {\bf I}
\ee
with $r$ positive and ${\bf I}$ unitary.
Finally we get
\be
 -\dot{q_i} - \Phi_{[ij]'j} =
    \frac{1}{m}\Big( L_{ik}q_k + M_{i[kl]} \Phi_{[kl]}
    + E_{i[jkl]} \Psi_{[jkl]} - \frac{\partial^2 s_{0}}{\partial e^2}e_{'i} \Big),
\label{tr1}\\ \ee
\be
 -\dot \Phi_{[ij]} - \Psi_{[ijk]'k} =
  \frac{1}{n}\Big( P_{[ij]p} q_p + Q_{[ij][kl]} \Phi_{[kl]}
  + R_{[ij][klm]} \Psi_{[klm]} \Big) +  \frac{m}{n}  q_{[i'j]}
\label{tr2}\\ \ee
\be
 -\dot \Psi_{[ijk]} =  \frac{1}{r}\Big( S_{[ijk]l} q_l + T_{[ijk][lm]}
    \Phi_{[lm]} + Z_{[ijk][lmn]} \Psi_{[lmn]} \Big) +
    \frac{n}{r}\Phi_{[ij]'k}. \label{tr3}
\ee

Here ${\bf E}$, ${\bf R}$, ${\bf S}$, ${\bf T}$ and ${\bf Z}$ are
Onsagerian conductivity tensors defined on the thermodynamic state
space. In this way we get the 10 moments theory by \re{balinte},
\re{tr1} and \re{tr2} together with the closure relation \re{tr3}.
It is worth noticing that the previous procedure of the extension
of the basic fields can be continued analogously. That represents
a general method to obtain extended thermodynamic systems together
with suitable closure relations.

\section{Extended thermodynamics of second sound}
Second sound, i.e. thermal wave propagation, is a typical low
temperature phenomenon which can be observed, for instance, in
dielectric crystals such as Sodium Fluoride (NaF) and Bismuth (Bi)
\cite{JacWal71a,NarDyn72a}. Its appropriate description requires an
extension of the classical Fourier's theory leading to the paradox
of an infinite speed of propagation of thermal disturbances
\cite{Cat48a}. From the microscopic point of view heat transport
at low temperature is modelled through the phonon gas
hydrodynamics \cite{Pei55b,Rei73b}. Phonons are quasi-particles
which obey the Bose-Einstein statistics. In a solid crystal at low
temperature they form a rarefied gas whose kinetic equation can be
derived similarly to that of an ordinary gas. Phonons may interact
among themselves and with lattice imperfections through two
different types of
processes: \\
i) Normal-(N) processes, that conserve the phonon momentum;\\
ii) Resistive-(R) processes, in which the phonon momentum is not
conserved. \\
The frequencies $\nu_{N}$ and $\nu_{R}$ of normal and
resistive processes determine the characteristic relaxation times
$\tau_{N}=\frac{1}{\nu_{N}}$ and $\tau_{R}=\frac{1}{\nu_{R}}$.
Diffusive processes take over when there are many more R-processes
than N-processes. If instead there are only few R-processes and
many more N-processes, then a wave like energy transport may
occur. This is said second sound propagation. Finally, there is a
third mechanism of energy transport due to the presence of
ballistic phonons travelling through the crystal without any
interaction. The three different mechanisms of energy transport
described above can be represented at a different level of
approximation in the framework of Extended Thermodynamics. To
achieve this goal let us suppose the dynamic variables are
the first moment $p_{i}$, $(i=1,2,3)$ and  the
momentum flux $N_{[ij]}$, $(i,j=1,2,3)$. By phonon gas hydrodynamics
\cite{Rei73b} follows that the functions $p_{i}$ are connected to
the heat flux by the relation $q_{i}=c^2p_{i}$ where $c$ means the
Debye's phonon velocity. Moreover, from kinetic theory we apply the
interrelation of the traces of consecutive currents, which in our case is
expressed by the definition $N_{[ii]}:= e$. Therefore it is convenient
to decompose the momentum flux $N_{[ij]}$ into an isotropic part and
a deviatoric part according to the equation \cite{DreStr93a}
\be
\label{decomp}
 N_{[ij]}=\frac{1}{3}e \delta_{ij} + N_{<ij>},
\ee
where $N_{<ij>}$ is symmetric and traceless.
Due to the decomposition above the energy density coincides with
the trace of $N_{[ij]}$. As a consequence,
we have only nine independent thermodynamic variables (9-field model)
whose evolution is controlled by the balance laws \cite{DreStr93a}
\begin{eqnarray}
\dot{p_i} + N_{[ik]'k} &=& P_{i},\label{linoc11}\\
\dot N_{[ij]} + J_{[ij]k'k}&=& P_{[ij]}. \label{linoc22}
\end{eqnarray}
Here $P_{i}$ and $P_{[ij]}$ are suitable production terms while
$J_{[ij]k}$ represents the flux of $N_{[ij]}$, to be assigned
through a closure relation.
Under suitable
constitutive assumptions on $J_{[ij]k}$ and $P_{[ij]}$, the balance of energy
\be
\label{balinte1} \dot{e} + c^2 p_{i'i} =0
\ee
can be obtained  by taking  the trace of (\ref{linoc22}). Later on
we illustrate such a property in more details by reinspecting
two typical examples often used in the applications.\\
\medskip
\medskip

{\bf 4-field model} \\
\medskip
\medskip

This model holds true under the approximation $\tau_{N}=0$ and
describes second sound effects without taking care for
N-processes. It is controlled by the equations
\be
 \dot{e} + c^2 p_{i'i} =0,
\label{balinte2} \ee
\be
 \dot{p_i} + \frac{1}{3}e_{'i} = -\frac{1}{\tau_{R}}p_{i},
\label{linoc111}\\ \ee
If one assumes $\Phi_{[ij]}=c^2 \frac{1}{3}e \delta_{ij}$ then
the system  above is recovered by (\ref{linoc1}) and
(\ref{linoc2}). Moreover, if in (\ref{linoc1}) we assume the isotropy of the
constitutive equations
\be
 M_{ikl}=0,\,\,\,\, L_{ij}= \frac{m}{\tau_{R}}\delta_{ij},
\ee
then we get
\be
 \dot{p_i} + \Big(\frac{1}{3} -
    \frac{1}{mc^2}\frac{\partial^2 s_{0}}{\partial e^2}\Big) e_{'i}
    = -\frac{1}{\tau_{R}}p_{i},\label{linoc1111}\\
\ee
On the other hand, due to the high speed of phonons \Big($42
\times 10^4 cm  sec^{-1}$ in NaF \cite{CimFri05a}\Big), the constant
$mc^2$ is very big. Hence, the coefficient $\frac{1}{mc^2}
\frac{\partial^2 s_{0}}{\partial e^2}$ in (\ref{linoc1111}) can be
neglected whenever the absolute temperature $T =\Big(
\frac{\partial s}{\partial e}\Big)^{-1}$ has no jumps, i.e.
whenever the Lax conditions for shock wave formation \cite{CimFri05a,
CimOli04a} are not fulfilled. In such a case (\ref{linoc1111})
reduces exactly to (\ref{linoc111}). Further assuming isotropy and
some more we get
\be
P_{[ij]k}=0,\,\,\,\, Q_{[ij][kl]}=0,\,\,\, \frac{m}{n} = c^2,
\ee
and taking into account the decomposition
$p_{[i'j]}=\frac{1}{3}p_{k'k}\delta_{ij} + p_{<i'j>}$, with $p_{<i'j>}$
symmetric and traceless, equation (\ref{linoc2}) can be rewritten as  follows
\be
\label{linoc4}
\frac{1}{3} \dot e \delta_{ij} + \frac{1}{3}p_{k'k}\delta_{ij}
+c^2 p_{<i'j>} =0.
\ee
Finally, taking the trace  of (\ref{linoc4}) we get the balance of energy
(\ref{balinte2}).
\medskip
\medskip

{\bf 9-field model} \\
\medskip
\medskip

Such an approximation holds for small $\tau_{N}$ and arbitrary
$\tau_{R}$. It is able to describe second sound taking into
account the effects due to the propagation of ballistic phonons.
Again, the calculated values of wave speeds are qualitatively but
not quantitatively in agreement with the measured ones.
The resulting system of equations is
\be
 \dot{e} + c^2 p_{i'i} =0,
\label{balinte3} \ee
\be
 \dot{p_i} + \frac{1}{3}e_{'i} + N_{<ij>'j}
    = -\frac{1}{\tau_{R}}p_{i},
\label{linoc11111}\\ \ee
\be
 \dot N_{<ij>} + c^2\frac{2}{5}p_{<i'j>}=
    -\frac{1}{\tau}N_{<ij>},
\label{linoc222} \ee
where
\be
\frac{1}{\tau}= \frac{1}{\tau_{R}} + \frac{1}{\tau_{N}}
\ee
is the total collision frequency. Let us observe that equation
(\ref{linoc222}) is not in the balance form since there is no flux
of $N_{<ij>}$ inside. Hence, nothing prevents to regard it as a
closure relation and use (\ref{linoc1}) and (\ref{linoc2}) to
derive the governing system of equations. Then, let us make the
identification
\be
 \Phi_{[ij]}=\frac{c^2}{3} e \delta_{ij}+ c^2 N_{<ij>},
\ee
and let us assume isotropic constitutive equations
\be
 M_{i[kl]}= 0, \,\,\,
 L_{ij}=\frac{m}{\tau_{R}}\delta_{ij},\,\,
 P_{[ij]k} = 0,\,\,\,
 Q_{[ij][kl]}=-\frac{1}{\tau}\delta_{ij}\delta_{kl}
    +\frac{1}{\tau'}\delta_{ik}\delta_{jl}\,\,
 \frac{m}{n} = c^2.
\ee

Here $Q_{[ij][kl]}$ is a constant isotropic tensor. There are only
two material quantities because of the symmetries of $Q_{[ij][kl]}$.
Here we require that $\frac{1}{\tau} = \frac{1}{\tau'}$.
Then by (\ref{linoc1}) and (\ref{linoc2}) we get
\be
 \dot{p_i} + \Big(\frac{1}{3} -
 \frac{1}{mc^2}\frac{\partial^2 s_{0}}{\partial e^2}\Big)e_{'i} + N_{<ij>'j}
 = -\frac{1}{\tau_{R}}p_{i},
\label{linoc111111}\\ \ee
\be
 \frac{1}{3} \dot e \delta_{ij} +\dot N_{<ij>}
    + \frac{1}{3}p_{k'k} \delta_{ij}
    + c^2 p_{<i'j>} = -\frac{1}{\tau} N_{<ij>}.
\label{linoc2222} \ee
The trace of (\ref{linoc2222}) yields the balance of the energy (\ref{balinte3}).
As a consequence, due to (\ref{balinte3}) the last equation reduces to
\be
 \dot N_{<ij>} + c^2 p_{<i'j>} = -\frac{1}{\tau} N_{<ij>}.
\label{linoc22222} \ee
Also in this case in the absence of shocks (\ref{linoc111111}) reduces to
(\ref{linoc11111}) while (\ref{linoc22222}) differs from (\ref{linoc222})
only for the coefficient in front of $p_{<i'j>}$. Such a discrepancy is of
no concern at that level since the 9-field model gives only qualitative
agreement with the experiments \cite{DreStr93a}.

\section{Conclusions}

We reinspected the problem of thermal wave propagation and proved
that the balance structure can be recovered with  the Verh\'as form
of the entropy current \re{entcurr} if the thermodynamic inductivity
coefficients $m_{ij}$ and $n_{ijkl}$ are constant in the general
entropy function (\ref{Gyarent}). The results above prove that the
balance form of evolution equations is compatible with the
mathematical structure of classical IT. The balance structure was
obtained only for a subset of the independent thermodynamic
variables. In fact, the highest order fluxes are controlled by
ordinary differential equations, here regarded as closure relations.

For our further investigations we have assumed some interrelations
of the consecutive currents \re{decomp}, to be compatible to the
moment interpretation of the kinetic theory. With assuming
isotropic material we have arrived to the expected system of
equations of the 4 and 9 moment theory of rigid heat conductors,
thereby demonstrating the compatibility of the two approaches. Let
us observe the minor, but important additional assumptions (e.g.
$\tau=\tau'$).

Nowadays in ET the balance form of the evolution equations is a well
motivated assumption from the kinetic theory and the entropy current
is a derived quantity with a restrictive potential structure
\re{jcurrret}. Moreover, the derived restrictions can be exploited
predictively in different particular cases (e.g. for monatomic gases
\cite{MulRug98b}, p.52-60. We do not know related calculations for
rigid heat conductors). In ET one of the important problems is to
give a reasonable closure of the arising hierarchy.

However, the applications of the phenomenological theory  in solids
and fluids (see e.g. \cite{Ver97b}) and recent generalizations of
the principles to incorporate nonlocality are far beyond the
validity of the kinetic theory. That was our most important
motivation of the recent investigations. In this paper we have shown
that with suitable constitutive assumptions on the form of the
entropy current and on the entropy one can recover the balance
structure. We have demonstrated that one can give a kind of natural
closure of the hierarchy.

On the other hand, the conditions  of using all the currents as
state variables and to restrict ourselves to constant inductivities
looks like too restrictive. However, to assume balance form
evolution equations, where the current is a constitutive function is
a weakly nonlocal theory from our point of view. E.g.
$$
\dot{q_i} = -\Phi_{ik,k}(e,q_i) = f(e,q_i,e_{'i},q_{i'j}).
$$
Calculation of the derivatives in the divergence of $\Phi$ gives
that the form of the evolution equation, defined by the function
$f$, depends on the space derivatives of the state variables, too.
Therefore, for a systematic investigation of the interrelation of
the evolution equations and the entropy current without imposing the
balance structure one should enlarge the basic thermodynamic space.
Hence, one should weaken an other basic assumption in ET, the
locality, by including the gradients of the state variables in the
constitutive state space \cite{Van03a,Van05a}. A study of the
effects of nonlocality on the evolution of higher order fluxes is
developed in \cite{CimVan04m}.

Simplifying the arguments that we wanted to collide here we can say
that although from the point of view of kinetic theory the specific
form of the entropy current is a restriction, from a
phenomenological point of view the balance structure looks like too
restrictive.

\section{Acknowledgements}

Thanks given for prof. Verh\'as for his valuable comments. This
research was supported by OTKA grants T034715, T034603 and T048489
by Cofin 2002 Modelli Matematici per la Scienza dei Materiali. A
grant of University of Basilicata is acknowledged as well.

\end{document}